\RequirePackage{ifpdf}
\ifpdf 
\documentclass[pdftex]{sigma}
\else
\documentclass{sigma}
\fi

\newcommand{\msp}{{\mspace{1mu}}}
\newcommand{\ep}[2]{{\epsilon_{#1#2}}}
\newcommand{\vep}{{\varepsilon}}
\newcommand{\ga}[3]{{\Gamma^{#1}{}_{#2#3}}}
\newcommand{\dg}{{\sqrt{|g|}\,}}
\newcommand{\bdt}{^{{\displaystyle\boldsymbol{\cdot}}}}

\newcommand{\bu}{{\boldsymbol u}}
\newcommand{\bw}{{\boldsymbol {u'}}}
\newcommand{\bp}{{\boldsymbol \pi}}
\newcommand{\bq}{{\boldsymbol \pi^{(1)}}}
\newcommand{\bpr}{{{\boldsymbol'}}}
\newcommand{\pr}[2]{{\boldsymbol{#1\!\cdot\!#2}}}

\newcommand{\nbu}{{\left\|\boldsymbol{u}\right\|}}
\newcommand{\nw}[2]{{\left\|{#1}\wedge{#2}\right\|}}
\newcommand{\bcdot}{{\boldsymbol{\cdot}}}

\numberwithin{equation}{section}

\begin{document}

\allowdisplaybreaks

\renewcommand{\PaperNumber}{016}

\FirstPageHeading

\renewcommand{\thefootnote}{$\star$}

\ShortArticleName{Variationality of Uniform Acceleration}

\ArticleName{The Variational Principle\\ for the Uniform
Acceleration and Quasi-Spin\\ in Two Dimensional
Space-Time\footnote{This paper is a contribution to the
Proceedings of the Seventh International Conference ``Symmetry in
Nonlinear Mathematical Physics'' (June 24--30, 2007, Kyiv,
Ukraine). The full collection is available at
\href{http://www.emis.de/journals/SIGMA/symmetry2007.html}{http://www.emis.de/journals/SIGMA/symmetry2007.html}}}

\Author{Roman Ya. MATSYUK}

\AuthorNameForHeading{R.Ya.~Matsyuk}

\Address{Institute for Applied Problems in Mechanics and
Mathematics, 15~Dudayev Str., L'viv, Ukraine}
\Email{\href{mailto:matsyuk@lms.lviv.ua}{matsyuk@lms.lviv.ua}}

\ArticleDates{Received October 31, 2007, in f\/inal form January
18, 2008; Published online February 06, 2008}

\Abstract{The variational principle and the corresponding
dif\/ferential equation
         for geodesic circles in two dimensional (pseudo)-Riemannian
        space are being discovered. The relationship with the physical notion of
        uniformly accelerated relativistic particle is emphasized. The known form
        of spin-curvature interaction emerges due to the presence
        of second order derivatives in the expression for the Lagrange function.
        The variational equation itself reduces to the unique invariant
        variational equation of constant Frenet curvature in two
        dimensional (pseudo)-Euclidean geometry.}

\Keywords{covariant Ostrohrads'kyj mechanics; spin; concircular
geometry; uniform accele\-ration}

\Classification{53A40; 70H50; 49N45; 83C10}

\section{Introduction}
It turns out that the notion of the uniformly accelerated
relativistic test particle world line~\cite{matsyuk:Hill81}
coincides with the notion of the geodesic circle in
pseudo-Riemannian geometry~\cite{matsyuk:Yano131}. In two
dimensions both these notions may be equivalently replaced by the
only condition that the f\/irst (and the only) Frenet curvature
$k$ of the curve in the consideration keeps constant along this
curve (in physical terms, along the world line of the test
particle).

In the natural parametrization by the arc length $s$ (in physical
terms, by
 proper time) all the notions, mentioned above, amount to the following
third order dif\/ferential equation (cf.\
Appendix~\ref{matsyuk:app5}):
\begin{gather}\label{matsyuk:Abraham}
\dfrac{D^3x^l}{ds^3}+g_{mn}\dfrac{D^2x^m}{ds^2}\dfrac{D^2x^n}{ds^2}\dfrac{Dx^l}{ds}=0.
\end{gather}
The left hand side of this equation is known in the physical
literature under the name of the {\it Abraham vector.} This vector
is believed to adequately represent the notion of the relativistic
acceleration of test particle~\cite{matsyuk:Hill81}. Of course,
the notion of geodesic circles does not depend on any arbitrary
reparametrization~$s({\xi})$ of the independent variable along the
curve in~(\ref{matsyuk:Abraham}).

One may ask whether geodesic circles could be the extremals of
some variational problem. Since the
equation~(\ref{matsyuk:Abraham}) involves third order derivatives,
it is appropriate to speak of higher order variational calculus.
Rather trying in na\"ive way to f\/ind a Lagrange function for the
equation~(\ref{matsyuk:Abraham}) (or, better, an equivalent
equation), it may happen more ef\/f\/icient to set a general
problem of f\/inding all possible third order variational
equations in two-dimensional space which keep the Frenet
curvature~$k$ constant along their solutions. Of course, one may
hope to solve in reasonable fullness such an {\it inverse
problem\/} only if restricted to certain types of the space
geometry. The natural restriction with regard to
(pseudo)-Riemannian space should, of course, be that of its local
model~-- the (pseudo)-Euclidean space. Thus one comes to the
necessity of the formulation of the {\it invariant inverse
variational problem} possessing the f\/irst integral~$k$. This
problem in two dimensions received the ultimate answer
in~\cite{matsyuk:MathMet16} which is quoted below as
Proposition~\ref{matsyuk:Unique}. This proposition says that to
obtain geodesic circles, one should, in two dimensions,
necessarily handle the Frenet curvature itself as the integrand of
the variational functional.

The fashion of including Frenet curvatures into the variational
integrand in physics literature, initiated, in our opinion, by the
papers of {\it Plyushchay\/} and {\it Nesterenko},  counts 30
years by now. Still, sometimes it escapes the common knowledge
that such quantities as, say, Frenet curvatures, are always
constants of motion along the extremals of the variational
problems where the Lagrange functions depend on nothing but the
corresponding curvatures themselves~\cite{matsyuk:Arodz}. This
fact being noticed long ago~\cite{matsyuk:thesis} for the
(pseudo)-Euclidean case, we present it here in general framework
by means of rather trivial observation of
Proposition~\ref{matsyuk:Ham}. Unfortunately, only in two
dimensions the f\/irst Frenet curvature taken as the Lagrange
function produces, due to its linearity in second derivatives, a
{\it third order}\/ equation.

In Section~\ref{matsyuk:equation} we deduce the variational
equation from the variational functional
\[
\int{kd\xi+ds}
\]
in (pseudo)-Riemannian two-dimensional space and prove that the
set of its extremals includes {\it all geodesic circles.}

Also we discovered quite an interesting fact, in our opinion, that
the force on the particle, having the same form as that produced
by spin-curvature interaction, emerges as a simple result of
calculating the variation of second derivatives in the variational
integrand.

\section{Preliminaries}
\subsection{Parameter independence and Ostrohrads'kyj mechanics}
The right action of the prolonged group $GL_{(2)}(\mathbb
R)\overset{\mathrm{def}}=\tilde J^2{}_0(\mathbb R,\mathbb R)_0$ of
parameter transformations (invertible transformations of the
independent variable $\xi$) on the second order velocities space
$T^2M=\{x^n,u^n,\dot u^n\}$ gives rise to the so-called
fundamental f\/ields on $T^2M$~(cf.\
Appendix~\ref{matsyuk:ParIndep}):
\begin{gather}\label{matsyuk:FundFields}
\zeta_1=u^n\dfrac{\partial}{\partial u^n}+2\dot
u^n\dfrac{\partial}{\partial\dot u^n},\qquad
\zeta_2=u^n\dfrac{\partial}{\partial\dot u^n}.
\end{gather}

A function $f$ def\/ined on $T^2M$ does not depend on the change
of independent variable $\xi$ (so-called parameter-independence)
if and only if
\begin{gather}\label{matsyuk:parind}
\zeta_1f=0,\qquad\zeta_2f=0.
\end{gather}

On the other hand, a function $L$ def\/ined on $T^2M$ constitutes
a parameter-independent variational problem with the functional $
\int L(x^n,u^n,\dot u^n)d\xi $
 if and only if the following Zermelo conditions are satisf\/ied~\cite[formula~(8.19)]{matsyuk:Logan}:
\begin{gather}\label{matsyuk:Zermelo}
\zeta_1L=L,\qquad\zeta_2L=0.
\end{gather}

\begin{note}
At this point it worths mentioning the dif\/ference between the
conditions (\ref{matsyuk:parind}) and (\ref{matsyuk:Zermelo}). It
would be instructive not to confuse the idea of the parameter
independence of a variational problem with that of the parameter
independence of Lagrange function. The Lagrange function of a
parameter-independent variational problem in not
parameter-independent itself.

The formulation of Zermelo conditions in the most general case of
arbitrary order and of arbit\-rary number of independent variables
(Zermelo--G\'eh\'eniau) may be read from~\cite{matsyuk:MKaw}
whereas the corresponding proof may be found in
\cite{matsyuk:DGA8} as well as in~\cite{matsyuk:thesis}.
\end{note}

Let us introduce the generalized momenta:
\begin{gather}\label{matsyuk:p}
p^{(1)}_n=\dfrac{\partial L}{\partial\dot u^n},\qquad
p_n=\dfrac{\partial L}{\partial u^n}-\frac{dp^{(1)}_n}{d\xi} .
\end{gather}
The Euler--Poisson equation is given by
\begin{gather}\label{matsyuk:E-P}
\mathcal E_n=\dfrac{\partial L}{\partial x^n}-\dfrac{d
p_n}{d\xi}=0 .
\end{gather}
The Hamilton function is given by:
\[
H=p^{(1)}_n\dot u^n+p_nu^n-L .
\]

\begin{lemma}[\cite{matsyuk:thesis,matsyuk:de Leon}\label{matsyuk:ham}]
\[
H=\zeta_1L-\frac{d\zeta_2L}{d\xi}-L .
\]
\end{lemma}

\begin{proposition}\label{matsyuk:Ham}
If a function $L_{\mathrm{II}}$ is parameter-independent and if a
function $L_{\mathrm I}$ constitutes a~pa\-ra\-me\-ter-independent
variational problem, then
 $L_{\mathrm{II}}$ is constant along the extremals of $L=L_{\mathrm{II}}+L_{\mathrm{I}}$.
\end{proposition}

\begin{proof}
By Lemma~\ref{matsyuk:ham} and in force of the
properties~(\ref{matsyuk:parind}) and~(\ref{matsyuk:Zermelo}) we
calculate
$H_{L_{\mathrm{II}}+L_{\mathrm{I}}}=\zeta_1(L_{\mathrm{II}}+L_{\mathrm{I}})
 -\frac{d}{d\xi}\zeta_2(L_{\mathrm{II}}+L_{\mathrm{I}})-L=-L_{\mathrm{II}}$.
But as far as the Hamilton function is constant of motion, so is
the~$L_{\mathrm{II}}$.
\end{proof}
\subsection{Inverse variational problem for geodesic circles}
As stated in the previous papers by the author, the general
third-order variational equation takes on the
shape~\cite{matsyuk:MathMet13,matsyuk:DAN}:
\begin{gather}\label{matsyuk:hamspin5}
A_{mn}\ddot u^n + \dot u^l\partial_{u^l}\, A_{mn}\dot u^n +
B_{mn}\dot u^n + c_m = 0 ,
\end{gather}
where the skew-symmetric matrix $A$, the matrix $B$, and
a column $c$ all depend on~$\xi$,~$x^n$, and~$u^n$, and satisfy
the following system of partial dif\/ferential equations:
\begin{equation}\label{matsyuk:hamspin6}
\begin{gathered}
       \partial_{_{_{_{{{\scriptstyle u}}}}}}{\!}_{[m}{}{{A}}_{nk]}=0,
\\
       2\,{{B}}_{[mn]}-3\,{\bf D_{_{\boldsymbol  1}}}{\kern.01667em}{{A}}_{mn}=0,
\\
       2\,\partial_{_{_{_{{{\scriptstyle u}}}}}}{\!}_{[m}{}{{B}}_{n]\,k}
               -4\,\partial_{_{_{_{{{\scriptstyle x}}}}}}{\!}_{[m}{}{{A}}_{n]\,k}
               +{\partial_{_{_{_{{{\scriptstyle x}}}}}}{\!}_{k}}{\,}{{A}}_{mn}
   +2\,{\bf D_{_{\boldsymbol  1}}}{\kern.01667em}{\partial_{_{_{_{{{\scriptstyle u}}}}}}{\!}_{k}}{\,}{{A}}_{mn}=0,
\\
        {\partial_{_{_{_{{{\scriptstyle u}}}}}}{\!}_{(m}}{}{{c}}_{n)}
               -{\bf D_{_{\boldsymbol  1}}}{\kern.01667em}{{B}}_{(mn)}=0,
\\
        2\,{\partial_{_{_{_{{{\scriptstyle u}}}}}}{\!}_{k}}{\,}\partial_{_{_{_{{{\scriptstyle u}}}}}}{\!}_{[m}{}{{c}}_{n]}
           -4\,\partial_{_{_{_{{{\scriptstyle x}}}}}}{\!}_{[m}{}{{B}}_{n]\,k}
           +{{\bf D_{_{\boldsymbol  1}}}}^{2}{\,}{\partial_{_{_{_{{{\scriptstyle u}}}}}}{\!}_{k}}{\,}{{A}}_{mn}
 +6\,{\bf D_{_{\boldsymbol  1}}}{\kern.0334em}\partial_{_{_{_{{{\scriptstyle x}}}}}}{\!}_{[m}{}{{A}}_{nk]}=0,
\\
       4\,\partial_{_{_{_{{{\scriptstyle x}}}}}}{\!}_{[m}{}{{c}}_{n]} -2\,{\bf D_{_{\boldsymbol
           1}}}{\kern.0334em}\partial_{_{_{_{{{\scriptstyle u}}}}}}{\!}_{[m}{}{{c}}_{n]} -{{\bf D_{_{\boldsymbol  1}}}}^{3}{\,}{{A}}_{mn}=0 .
\end{gathered}
\end{equation}
Here the dif\/ferential operator $\bf D_{_{\boldsymbol  1}}$ is
the lowest order truncated operator of total derivative, ${\bf
D_{_{\boldsymbol  1}}}=\partial_{\xi}+u^l\partial_{x^l}$. We refer
to Appendix~\ref{matsyuk:Inv} on some additional comments
concerning the nature of the variationality
conditions~(\ref{matsyuk:hamspin6}).

On the other hand, we are interested only in dif\/ferential
equations which enjoy the (pseudo)-Euclidean symmetry. Because of
that in two dimensions the skew-symmetric matrix $A$
in~(\ref{matsyuk:hamspin5}) is invertible, there is no problem to
formulate the concept of the symmetry in the way
\begin{gather}\label{matsyuk:symm}
\left.\big(X\,\mathcal E_n\big)\right|_{\mathcal E=0}=0 ,
\end{gather}
where~$\mathcal E$ denotes the left hand side of the
equation~(\ref{matsyuk:hamspin5}) and~$X$ stands for the
inf\/initesimal generator of the (pseudo)-Euclidean group.

The system of partial dif\/ferential
equations~(\ref{matsyuk:hamspin6}) together
with~(\ref{matsyuk:symm}) was solved in~\cite{matsyuk:MathMet16}
and the following result was established:
\begin{proposition}\label{matsyuk:Unique}
Let some system of third order differential
equations~\eqref{matsyuk:hamspin5} enjoy the following properties:
\begin{itemize}\itemsep=0pt
\item the conditions~\eqref{matsyuk:hamspin6} are satisfied; \item
the system~\eqref{matsyuk:hamspin5} possesses Euclidean symmetry
according to~\eqref{matsyuk:symm}; \item the Euclidean geodesics
$\boldsymbol {\dot u}=\boldsymbol0$ enter in the set of solutions
of~\eqref{matsyuk:hamspin5}; \item $\dfrac{dk}{d\xi}=0$ along the
solutions of~\eqref{matsyuk:hamspin5}.
\end{itemize}
Then
\begin{gather}\label{matsyuk:Eps}
\mathcal E_n =\frac{\ep n l\ddot
u\msp^l}{\nbu^3}-3\,\frac{(\pr{\dot u}{u})}{\nbu^5}\,\ep n l\dot
u\msp^l + m\,\frac{(\pr u u)\dot u_n-(\pr{\dot u}{u})u_n}{\nbu^3}
.
\end{gather}
\end{proposition}

In the above statement the denotation~$\ep n m$ stands for the
skew-symmetric Levi-Civita symbol. This
expression~(\ref{matsyuk:Eps}) may be obtained as an
Euler--Poisson expression for
 the Lagrange function
\begin{gather}\label{matsyuk:flat}
L=\dfrac{e_{mn} u^m \dot u^n}{\nbu^3}-m\,\nbu .
\end{gather}
The f\/irst addend in~(\ref{matsyuk:flat}) is sometimes called
{\em the signed Frenet curvature\/}~\cite{matsyuk:Arreaga} in
$\mathbb E^2$. This, along with the observation that in two
dimensional (pseudo)-Riemannian space the Frenet curvature
\begin{gather}\label{matsyuk:Frenet}
k=\dfrac{\nw{\bu}{\bw}}{\nbu^3}=\pm\dg\,\dfrac{\ep k
nu^ku'\msp^n}{\nbu^3}
\end{gather}
 depends linearly on the {\it covariant derivative} $\bw$ and
thus produces at most the third order Euler--Poisson equation,
suggests the next assertion, based on
Proposition~\ref{matsyuk:Ham}:
\begin{proposition}[\cite{matsyuk:Boundary Problems,matsyuk:MathMet16}]
The variational functional with the Lagrange function
\begin{gather}\label{matsyuk:L total}
L^{\mathcal R}=\dg\,\dfrac{\ep k nu^ku'\msp^n}{\nbu^3}-m\,\nbu
\end{gather}
produces geodesic circles in two dimensional Riemannian space.
\end{proposition}
\begin{note}
The Lagrange function~(\ref{matsyuk:flat}) looks like a
relativistic analogue of the one recently treated
in~\cite{matsyuk:Acatrinei}.
\end{note}

\section{The variational equation for geodesic circles}
\subsection{The generalized covariant momenta}
It is known that the Euler--Poisson expression (in this paper, --
of the third order)~(\ref{matsyuk:E-P}) constitutes a covariant
geometric object. But the conventional momentum~$p_n$
{from}~(\ref{matsyuk:p})
 does not.
Therefore we introduce covariant momenta $\bp$ and $\bq$ that, in
the case of the Lagrange function we shall deal with, represent
some relative vectors in the following manner. First, we assume
that the Lagrange function depends on the variables $\dot u^n$
exclusively through the covariant derivative~$\bw$.  Moreover, in
the case handled in this article, the partial
derivatives~$\dfrac{\partial L}{\partial \dot u^n}$
and~$\dfrac{\partial L}{\partial u'\msp^n}$ coincide. Let us
introduce the ``truncated'' partial derivatives
\begin{gather}\label{matsyuk:u-truncated}
\dfrac{\overline{\strut\partial L}}{\partial u^n}=\dfrac{\partial
L}{\partial u^n}-\dfrac{\partial u'\msp^q}{\partial
u^n}\dfrac{\partial L}{\partial u'\msp^q},
\end{gather}
and, for future use,
\begin{gather}
\label{matsyuk:x-truncated} \dfrac{\overline{\strut\partial
L}}{\partial x^n}=\dfrac{{\partial L}}{\partial
x^n}-\dfrac{\partial u'\msp^q}{\partial x^n}\dfrac{\partial
L}{\partial u'\msp^q} .
\end{gather}
Then the quantities
\begin{gather}
 \bq =\dfrac{\partial L}{\partial \bw}\qquad \text{and} \label{matsyuk:pi1}\\
 \bp =\dfrac{\overline{\strut\partial L}}{\partial \bu}-\bq\bpr \qquad \text{(see~(\ref{matsyuk:u-truncated}))} \label{matsyuk:pi-1}\\
\phantom{\bp}{} =\dfrac{\partial L}{\partial u^n}-2\,\ga q m n u^m
\pi^{(1)}{}_q
 -{\pi^{(1)}}\msp'{}_n \qquad \text{(by virtue of~(\ref{matsyuk:AppCov}))}
 \label{matsyuk:pi}
\end{gather}
may each be applied the operation of the covariant
dif\/ferentiation according to the
formulae~(\ref{matsyuk:AppCov}). For example, the covariant
momentum~$\bq$ produces its covariant derivative
\begin{gather}\label{matsyuk:pi1'}
{\pi^{(1)}}\msp'{}_n=\dot\pi^{(1)}{}_n-\ga m l n \pi^{(1)}{}_m u^l
.
\end{gather}
In the above terms the quantity~$p_n$ of~(\ref{matsyuk:p}) is
expressed as follows:
\begin{gather}
 p_n=\pi_n+2\,\ga q m n u^m\pi^{(1)}{}_q+\pi^{(1)}\msp'{}_n
 \qquad\text{(on base of (\ref{matsyuk:pi}))} \nonumber\\
\phantom{p_n=}{}-\dot\pi^{(1)}{}_n
 \qquad\text{(on base of (\ref{matsyuk:pi1}))} \nonumber \\
\phantom{p_n}{} =\pi_n+\ga q m n u^m\pi^{(1)}_q \qquad\text{(on
base of (\ref{matsyuk:pi1'})).}\label{matsyuk:p_n}
\end{gather}
Dif\/ferentiating (\ref{matsyuk:p_n}) and applying
(\ref{matsyuk:AppCov}) to express the ordinary derivatives of
$\pi_n$ and $u^n$ in terms of $\bp\bpr$ and $\bu\bpr$ along
with~(\ref{matsyuk:pi1'}), gives:
\begin{gather*}
\dot p_n=\big(\pi'{}_n+\ga l m n \pi_l u^m\big)+\dfrac{\partial
\ga l m n}{\partial x^k}
 u^ku^m\pi^{(1)}{}_l \\
\phantom{\dot p_n=}{} +\big(\ga l m n {u'}\msp^m-\ga l m n
 \ga m q k u^qu^k\big)\pi^{(1)}{}_l
 +\ga l m n u^m\big({\pi^{(1)}}\msp'{}_l
 +\ga q k l \pi^{(1)}{}_q u^k\big)  \\
\phantom{\dot p_n}{}        =\pi'{}_n+\big(\pi^{(1)}\msp'{}_l+\pi_l\big)\ga l m n u^m+\pi^{(1)}{}_l\ga l m n u'\msp^m \\
\phantom{\dot p_n=}{}        +\pi^{(1)}{}_q u^m u^k\left(\ga l m n
\ga q l k+\dfrac{\partial \ga q m n}{\partial x^k}
        -\ga q l n \ga l m k\right).
\end{gather*}
Now the Euler--Poisson expression~(\ref{matsyuk:E-P}) may be
handled with the help of our ``truncated''
$x$-de\-rivative~(\ref{matsyuk:x-truncated}) in the following way:
\begin{gather}\label{matsyuk:E II}
\mathcal E_n=\dfrac{\overline{\strut\partial L}}{\partial
x^n}-\pi'{}_n
        -\big(\pi^{(1)}\msp'{}_l+\pi_l\big)\ga l m n u^m
        -\pi^{(1)}{}_l\ga l m n u'\msp^m
        -\pi^{(1)}{}_l u^mu^kR_{nkm}{}^l ,
\end{gather}
making use of the def\/inition of the curvature
tensor~(\ref{matsyuk:AppR}).

\subsection[The Euler-Poisson expression for the signed Frenet curvature]{The Euler--Poisson expression for the signed Frenet curvature}\label{matsyuk:equation}

Let us carry out the previous considerations taking for the
Lagrange function $L_{\rm II}$ the signed Frenet curvature:
\begin{gather}\label{matsyuk:L II}
L_{\rm II}=\dg\,\dfrac{\ep m nu^mu'\msp^n}{\nbu^3} .
\end{gather}
Making use of (\ref{matsyuk:AppDxg}) and (\ref{matsyuk:AppDxG})
one gets immediately
\begin{gather}\label{matsyuk:DL bar}
\dfrac{\overline{\strut\partial L_{\rm II}}}{\partial
x^n}=\dg\,\dfrac{\ga q q n
        \ep m ku^mu'\msp^k}{\nbu^3}
        -3\dg\,\dfrac{\ga l n mu_lu^m}{\nbu^5} .
\end{gather}
The expression for $\bq$ is obtained {from}~(\ref{matsyuk:pi1}) in
still easier way:
\begin{gather}\label{matsyuk:Pi1}
\pi^{(1)}{}_n=\dg\,\dfrac{\ep k nu^k}{\nbu^3} .
\end{gather}
The covariant derivative of~$\bq$ is presented by the
decomposition formula~(\ref{matsyuk:pi1'}). While calculating
$\dot\pi^{(1)}{}_n$ in~(\ref{matsyuk:pi1'}) we prof\/it both
{from} (\ref{matsyuk:AppDxg}) and {from}~(\ref{matsyuk:AppModu'}).
Then we replace $\pi^{(1)}{}_n$ in~(\ref{matsyuk:pi1'}) with
(\ref{matsyuk:Pi1}) to obtain:
\begin{gather}
\pi^{(1)}\msp'{}_n=
         \dg\,\dfrac{\ga q q lu^l\ep k nu^k}{\nbu^3}
        +\dg\,\dfrac{\ep k n}{\nbu^3}
        \big(u'\msp^k-\ga k m lu^lu^m\big)\nonumber\\
\phantom{\pi^{(1)}\msp'{}_n=}{}
        -3\dg\,\dfrac{(\pr u{u'})\ep k nu^k}{\nbu^5}
        -\dg\,\dfrac{\ga l n m\ep k lu^ku^m}{\nbu^3}.\label{matsyuk:Pi1'-1}
\end{gather}
Finally, the simplif\/ication formula~(\ref{matsyuk:g-simpl}) may
be used to produce
\begin{gather}\label{matsyuk:Pi1'}
        \pi^{(1)}\msp'{}_n=
        \dg\,\dfrac{\ep k nu'\msp^k}{\nbu^3}
        -3\dg\,\dfrac{(\pr u{u'})\ep k nu^k}{\nbu^5} .
\end{gather}
The calculation of the sum $\bp+\bq\bpr$ on grounds of
(\ref{matsyuk:pi-1}) is straightforward:
\[
\pi_n+\pi^{(1)}\msp'{}_n=-\dg\,\dfrac{\ep k nu'\msp^k}{\nbu^3}
        -3\dg\,\dfrac{u_n\ep l k u^lu'\msp^k}{\nbu^5} .
\]
Or, alternatively, utilizing~(\ref{matsyuk:e-simpl}), one obtains
\begin{gather}\label{matsyuk:Pi+Pi1'}
\pi_n+\pi^{(1)}\msp'{}_n=2\dg\,\dfrac{\ep k nu'\msp^k}{\nbu^3}
        -3\dg\,\dfrac{(\pr u{u'})\;\ep k nu^k}{\nbu^5} .
\end{gather}
To calculate~$\bp$, we extract~(\ref{matsyuk:Pi1'})
{from}~(\ref{matsyuk:Pi+Pi1'}):
\begin{equation}\label{matsyuk:pi_n}
\pi_n=\dg\,\dfrac{\ep k nu'\msp^k}{\nbu^3} .
\end{equation}
Now, the Euler--Poisson expression~(\ref{matsyuk:E II}) for the
Lagrange function~(\ref{matsyuk:L II}) by virtue of
(\ref{matsyuk:DL bar}), (\ref{matsyuk:Pi1}),
and~(\ref{matsyuk:Pi+Pi1'}) reduces to
\begin{gather}\label{matsyuk:E II ultimate}
\mathcal E_n(L_{\rm II})=-\pi'{}_n-\pi^{(1)}{}_qR_{nkm}{}^qu^mu^k
,
\end{gather}
where the covariant derivative of the momentum~$\bp$ may
be calculated starting with the definiton~(\ref{matsyuk:AppCov}) applied 
to~(\ref{matsyuk:pi_n}), 
wherein~$(u'\msp^k)\bdt$ should be calculated again along the lines of~(\ref{matsyuk:AppCov}): 
\[
(u'\msp^k)\bdt=u''\msp^k-\ga k l mu'\msp^mu^l.
\]
Using then (\ref{matsyuk:AppDxg}) and (\ref{matsyuk:AppModu'}) along with the 
simplification formula~(\ref{matsyuk:g-simpl}) once more, we calculate 
\[
\pi'_n=-3\,\dfrac{\dg(\pr u{u'})}{\nbu^5}\,\ep k nu'\msp^k
        +\dfrac{\dg}{\nbu^3}\,\ep k nu''\msp^k.
\]
Thus in the
disclosed form the Euler--Poisson expression for the total
Lagrange function~(\ref{matsyuk:L total}) reads:
\begin{gather}
        \mathcal E^{\mathcal R}_n=
        -\dg\,\dfrac{\ep k nu''\msp^k}{\nbu^3}
        +3\dg\,\dfrac{(\pr u{u'})\ep k nu'\msp^k}{\nbu^5}\nonumber\\
\phantom{\mathcal E^{\mathcal R}_n=}{}
        +m\,\dfrac{(\pr u u)\,u'{}_n-(\pr{u'}u)\,u_n}{\nbu^3}
        -\dfrac{\dg}{\nbu^3}\,\ep m qR_{nkl}{}^qu^mu^lu^k .\label{matsyuk:E alt-1}
\end{gather}
The Euler--Poisson equation, solved with respect to highest
derivative, now is
\begin{gather}\label{matsyuk:E alt}
        \dg\left(\dfrac{u''\msp^k}{\nbu^3}
        -3\,\dfrac{(\pr u{u'})u'\msp^k}{\nbu^5}\right)
        -m\,\dfrac{(\pr u u)\,e^{nk}u'{}_n-(\pr{u'}u)\,e^{nk}u_n}{\nbu^3}
        +e^{nk}\mathcal R_n=0 ,
\end{gather}
where we used the notion of the contravariant skew-symmetric
Levi-Civita symbol~$e^{mn}=\det[g_{pq}]\,g^{mk}g^{nl}\ep k l$, and
introduced the shortcut
\begin{gather}\label{matsyuk:Rn}
\mathcal R_n=\frac{\dg}{\nbu^3}\,\ep m qR_{nkl}{}^qu^mu^lu^k
\end{gather}
 to denote the force on the particle, evoked by the curvature of the (pseudo)-Riemannian structure.
 Equation~(\ref{matsyuk:E alt-1}) obviously generalizes that of~(\ref{matsyuk:Eps}).
\begin{note}
If, on the other hand, we had introduced a {\it spin
tensor\/}~$S_{mq}=\pi^{(1)}{}_mu_q-\pi^{(1)}{}_qu_m$, the second
term on the right of formula~(\ref{matsyuk:E II ultimate}) would
have got an interpretation as
 the force, evoked by the existence of quasi-classical spin, and would
coincide with that present in Dixon's
equation~\cite{matsyuk:Dixon}.
\end{note}
\begin{proposition}
The differential equation~\eqref{matsyuk:E alt} describes  all
geodesic circles.
\end{proposition}

\begin{proof}
We have to show that, for each solution of~(\ref{matsyuk:Abraham})
it is always possible to perform such transformation of the
independent variable~$s$ to~$\xi$ that after that change of
variables the equation~(\ref{matsyuk:E alt}) will hold.

First, let us rewrite the dif\/ferential equation of geodesic
circles (\ref{matsyuk:Abraham}) in arbitrary
pa\-ra\-met\-rization:
\begin{gather}\label{matsyuk:Abraham arbitrary}
\dfrac{\bw\bpr}{\nbu^3}=\dfrac{\pr
u{u''}}{\nbu^5}\,\bu+3\,\dfrac{(\pr u{u'})}{\nbu^5}\,\bw
        -3\,\dfrac{(\pr u{u'})^2}{\nbu^7}\,\bu .
\end{gather}
We add to it one more equation, which will play the role of such a
one, that f\/ixes the parametrization along the curve:
\begin{gather}\label{matsyuk:Fix par}
\dg\left(\dfrac{\pr u{u''}}{\nbu^3}-3\,\dfrac{(\pr
u{u'})^2}{\nbu^5}\right)=
e^{nk}\left(\dfrac{m}{\nbu}\,u'{}_nu_k-u_k\mathcal R_n\right).
\end{gather}
This equation is consistent with the equation~(\ref{matsyuk:E
alt}) because it presents nothing more than a mere consequence
of~(\ref{matsyuk:E alt}), obtained by means of the contraction
with~$u_k$. Further, let us substitute~$\dfrac{u''\msp^k}{\nbu^3}$
in~(\ref{matsyuk:E alt}) from the equation~(\ref{matsyuk:Abraham
arbitrary}):
\begin{gather}\label{matsyuk:**}
-\dg\left(\dfrac{\pr u{u''}}{\nbu^5}-3\,\dfrac{(\pr
u{u'})^2}{\nbu^7}\right)u^k= e^{kn}\left(m\,\dfrac{(\pr u
u)\,u'{}_n+(\pr {u'}u)\,u_n}{\nbu^3}-\mathcal R_n\right) .
\end{gather}
It remains to insert the left hand side of~(\ref{matsyuk:Fix par})
into the left hand side of~(\ref{matsyuk:**}) and to notice that
in two-dimensional case what comes out is the identity:
\begin{gather}\label{matsyuk:identity}
\dfrac{1}{\pr
uu}\,\left(\dfrac{m}{\nbu}e^{nl}u\msp'{}_nu_l-e^{nl}u_l\mathcal
R_n\right)\,u^k \equiv
e^{kn}\,\left(\dfrac{m}{\nbu}\,u\msp'{}_n-m\,\dfrac{\pr{u'}u}{\nbu^3}\;u_n-\mathcal
R_n\right) .
\end{gather}
To see that~(\ref{matsyuk:identity}) is satisf\/ied identically,
it suf\/f\/ices to apply the simplif\/ication
formula~(\ref{matsyuk:e-simpl}) separately to the terms
 involving~$\mathcal R$ and then to what remains, and to notice that $u^n\mathcal R_n=0$
due to the previously introduced notation~(\ref{matsyuk:Rn}).
\end{proof}

\appendix
\section{Appendix}
\subsection[Parameter independence and fundamental fields]{Parameter independence and fundamental f\/ields}\label{matsyuk:ParIndep}

{\bf The $T^rM$: loose comments.} In the variational calculus
various objects depend not only on the variables of the
conf\/iguration space, but also on the velocities and on the
accelerations of the f\/irst as well as of higher orders. The
space of this extended number of variables may be introduced in
dif\/ferent ways, as it has in fact been by dif\/ferent authors.
In present paper we choose to use the def\/inition, belonging to
Ehresmann~\cite{matsyuk:Ehresmann}, of the higher order velocity
space $T^rM=J^r{}_0(\mathbb R,M)_0$ as the set of jets of mappings
from a neighborhood of the origin in $\mathbb R$ to the
conf\/iguration space $M={x^n}$, which all start at the origin
$0\in\mathbb R$. This space $T^rM$ is naturally endowed by several
geometric structures. The Reader may consult the monographs
\cite{matsyuk:YanoIshihara} and \cite{matsyuk:de Leon} on the
subject. For the purposes of this paper it suf\/f\/ices to think
about $T^rM$ as a manifold of the variables $x^n,  u^n ,\ldots,
u_{r-1}^n$, constructed of the successive derivatives of the
conf\/iguration space variables $x^n$ by the independent
evolutionary one.

\medskip
\noindent {\bf The fundamental f\/ields.} The group of invertible
jets $\tilde J^2{}_0(\mathbb R,\mathbb R)_0$ with the source as
well as the target at the origin $0\in\mathbb R$ presents an
appropriate geometrical concept when speaking about local
transformations of the parameter along the curve in a manifold.
Let again the variables $u^n=\dot x^n$, $\dot u^n=\ddot x^n$
denote the f\/irst and the second derivatives of coordinates along
the curve~$x^n(\xi)$, so that the jet $j^{(2)}_0x$ at zero is
presented by the array of second-order polynomials $u^n\xi+\dot
u^n\xi^2$. Another jet, $j^{(2)}\sigma\in\tilde J^2{}_0(\mathbb
R,\mathbb R)_0$, which is presented by the polynomial
$\alpha\xi+\frac 1 2 \beta\xi^2$, acts on the right upon the
previous one by the composition:
\begin{gather}
j^{(2)}_0x\cdot j^{(2)}\sigma =j^{(2)}(x\circ\sigma)
=u^n\cdot\left(\alpha\xi+\tfrac 1 2 \beta\xi^2\right)+\tfrac 1 2 \dot u^n\cdot(\alpha\xi+\tfrac 1 2 \beta\xi^2)^2 \nonumber\\
 \phantom{j^{(2)}_0x\cdot j^{(2)}\sigma }{} =u^n\alpha\xi+\tfrac 1 2 (u^n\beta+\dot u^n\alpha^2)\,\xi^2\quad \mod{o(\xi^2)}.\label{matsyuk:one*}
\end{gather}
Consider an $\vep$-shift $\sigma_{\vep}(\xi)$ of the
transformation $\sigma$ of the independent variable~$\xi$ in the
local curve expression~$x^n(\xi)$. This shift is presented in
$\tilde J^2{}_0(\mathbb R,\mathbb R)_0$ by some
$\alpha_{\vep}=\frac{\partial\sigma_{\mathstrut\vep}}{\strut\partial\xi}(0)$
and
$\beta_{\vep}=\frac{\partial^2\sigma_{\mathstrut\vep}}{\strut\partial\xi^{2}}(0)$.
It evokes a corresponding f\/low $\{u^n_{\vep},\,\dot
u^n_{\vep}\}\in T^2M$ induced by the action~(\ref{matsyuk:one*}):
\begin{gather*}
u^n_{\vep}=\alpha_\vep u^n ,\qquad \dot u^n_{\vep} =\beta_{\vep}
u^n+\alpha_{\vep}{}^2\dot u^n .
\end{gather*}
The generator of this f\/low is:
\begin{gather}\label{matsyuk:flow generator}
u^n\left.\frac{d\alpha_{\vep}}{d\vep}\right|_{\vep=0}\;\frac{\partial
}{\partial u^n} \;
        +\left.\left(\frac{d\beta_{\vep}}{d\vep}\,u^n
        +2\alpha_{\vep}\,\frac{d\alpha_{\vep}}{d\vep}\,\dot u^n\right)\right|_{\vep=0}\; \frac{\partial }{\partial \dot u^n}\,.
\end{gather}
Now it suf\/f\/ices to limit ourselves to the shift
$\sigma_{\vep}(\xi)$ of the form
$\sigma_{\vep}(\xi)=\xi+\vep\tau(\xi)$. Then the
generator~(\ref{matsyuk:flow generator}) reads:
\[
u^n\,\frac{d\tau}{d\xi}(0)\;\frac{\partial }{\partial u^n}
        +\left(u^n\,\frac{d^2\tau}{d\xi^2}(0)
        +2\dot u^n\,\frac{d\tau}{d\xi}(0)\right)  \frac{\partial }{\partial \dot u^n} .
\]
Taking $\tau(\xi)=\xi$ we obtain the f\/irst fundamental f\/ield
$\zeta_1$ in~(\ref{matsyuk:FundFields}). Choosing $\tau(\xi)=\frac
1 2 \xi^2$ we obtain the second fundamental f\/ield $\zeta_2$
in~(\ref{matsyuk:FundFields}).

\subsection{The inverse variational problem: loose comments}\label{matsyuk:Inv}

The system of equations~(\ref{matsyuk:hamspin6}) along with the  specif\/ic
guise~(\ref{matsyuk:hamspin5}) of the Euler--Poisson equation
arises as the general solution of the so-called {\it inverse
variational problem in the calculus of variations.} This problem
has been attacked by numerous authors from varying points of view
based on dif\/ferent approaches. Roughly speaking, the problem
consists in f\/inding out the criterion that for an a priori given
dif\/ferential equation  there locally exists a Lagrange function,
from  which this equation follows by the variational procedure,
applied to the corresponding action functional. As far as the act
of the variation of the functional may be expressed in the form of
some operator~$\delta$ action on the Lagrange function~$L$ in such
a way that $\mathcal E=\delta L$ be a~well-def\/ined geometric
object, represented in local coordinates by means of well-known
system of the Euler--Poisson expressions~${\mathcal E_n}$, it is
tempting to give such a~def\/inition of~$\delta$, wherewith the
cohomology complex property~$\delta^2$ should hold. The
equations~(\ref{matsyuk:hamspin6}) express the fact that the
dif\/ferential form~$\mathcal E_ndx^n$ is closed with respect
to~$\delta$. It is not our intention in this paper to discuss
further the ways of def\/ining the operator~$\delta$. Along with
the sources~\cite{matsyuk:MathMet13,matsyuk:DAN} we wish to show
the interested reader to the book by Olga
Krupkov\'a~\cite{matsyuk:OKrup} with the plentitude of references
therein. However, our explanation here was based on the approach
of Tulczyjew~\cite{matsyuk:Tulcz}.

\subsection[Simplifications of exterior products]{Simplif\/ications of exterior products}\label{matsyuk:AppCircles}

In two dimensions some vector and tensor skew-symmetric
expressions simplify drastically. Let~$a^n$,~$b^n$ and~$c^n$ denote
arbitrary vectors and let $\ga l m n$ denote for a moment an
arbitrary three-index quantity. The following two
simplif\/ications keep true if the underlying (pseudo)-Riemannian
 manifold is two-dimensional:
\begin{gather}\label{matsyuk:e-simpl}
        g_{mn}a^ma^n\ep l k b^l-g_{mn}a^mb^n\ep l k a^l+a_k\ep m n a^mb^n =0 ,\\
\label{matsyuk:g-simpl}
        \ep m n a^mb^n\ga l l kc^k - \ep m nb^n \ga m l k a^lc^k  
                + \ep m na^n \ga m l k b^lc^k=0 .
\end{gather}
The proof consists in the ingenuous calculation.

\subsection{Formulae from (pseudo)-Riemannian geometry}

Here we list some well-known relations, involving Christof\/fel
symbols, curvature tensor, and the covariant derivative. The
latter will be denoted by the prime superscript, while to denote
the ordinary derivative the dot will be used, e.g.~$u^n=\dot x^n$,
$\dot u^n=\ddot x^n$, etc.
\begin{gather}\label{matsyuk:AppCov}
        a'\,^n=\dot a^n+\ga n l m a^m u^l , \qquad a'{}_n=\dot a_n-\ga m l n a_m u^l ,\\
\label{matsyuk:AppDxG}
        \dfrac{\partial g_{mn}}{\partial x^k}=g_{ml}\ga l k n +g_{nl}\ga l k m ,\\
\label{matsyuk:AppDxg}
        \dfrac{\partial }{\partial x^n}\dg=\dg\ga l l n,\qquad\text{where}\quad g=\det[g_{nm}],\\
\label{matsyuk:AppModu'}
        \left(\dfrac{1}{\strut\nbu^3}\right)'=-3\,\dfrac{\bu\bcdot\bw}{\nbu^5} ,\\
\label{matsyuk:AppR}
        R_{kmn}{}^l=\dfrac{\partial \ga l k n}{\partial x^m}
        -\dfrac{\partial \ga l m n}{\partial x^k}+\ga l m q \ga q k n
        -\ga l k q \ga q m n .
\end{gather}

\subsection{Characterization of geodesic circles}\label{matsyuk:app5}

It is not dif\/f\/icult to check that the condition
$\frac{dk}{d\xi}=0$ in two dimensions is equal to the
equation~(\ref{matsyuk:Abraham}) on the shell $g_{mn}u^mu^n=1$.
More precisely, at the constraint manifold
\begin{gather}\label{matsyuk:Appu2} \bu\cdot\bu=1 ,\\
 \label{matsyuk:Appu2'} \bu\cdot\bw=0 ,\\
 \label{matsyuk:Appu2''} \bw\cdot\bw+\bu\cdot\bw\bpr=0 ,
\end{gather}
the dif\/ferential equation
\begin{gather}\label{matsyuk:Appk'} k^2\msp'\equiv\pm\,\pr{u'}{u''} =0
\end{gather}
is equivalent to the dif\/ferential equation of the geodesic
circles
\begin{gather}\label{matsyuk:AppGS}
\bw\bpr+(\pr{u'}{u'})\,\bu=\boldsymbol 0 .
\end{gather}

In fact, {from}~(\ref{matsyuk:Appk'}) and~(\ref{matsyuk:Appu2''})
one solves for $\bw\bpr$:
\[
u''{}_l=\dfrac{\ep l m u'\msp^m}{\ep m n u'\msp^m
u^n}\;\pr{u'}{u'} ,
\]
and then implementing~(\ref{matsyuk:e-simpl}) with the help of
(\ref{matsyuk:Appu2}) and (\ref{matsyuk:Appu2'})
obtains~(\ref{matsyuk:AppGS}). Conversely,
the~(\ref{matsyuk:Appk'}) in nothing but the~(\ref{matsyuk:AppGS})
contracted with~$\bw$ at the constraint
manifold~(\ref{matsyuk:Appu2'}).

\subsection*{Acknowledgements}

This work was supported by the Grant GA\v CR 201/06/0922 of Czech
Science Foundation.

\pdfbookmark[1]{References}{ref}
\LastPageEnding

\begin{thebibliography}{99}

\footnotesize\itemsep=0pt

\bibitem{matsyuk:Hill81}
        Hill E.L., On the kinematics of uniformly accelerated motions
        and classical magnetic theory, {\it Phys. Rev.} {\bf 72} (1947), 143--149.

\bibitem{matsyuk:Yano131}
        Yano K., Concircular geometry~I. Concircular transformations,
         {\it Proc. Imp. Acad. Jap.} {\bf 16} (1940), 195--200.

\bibitem{matsyuk:MathMet16}Matsyuk R.Ya., Variational principle for uniformly accelerated
        motion, {\it Mat. Metody Fiz.-Mekh. Polya} {\bf 16} (1982), 84--88 (in Russian).

\bibitem{matsyuk:Arodz}Arod\'z H., Sitarz A., W\c egrzyn P., On relativistic
        point particles with curvature-dependent actions,
        {\it Acta Phys. Polon.~B} {\bf 20} (1989), 921--939.

\bibitem{matsyuk:thesis}Matsyuk R.Ya., Poincar\'e-invariant equations of
        motion in Lagrangian mechanics with higher derivatives,
        PhD Thesis, Institute for Applied Problems in Mechanics and Mathematics, L'viv, 1984 (in Russian).

\bibitem{matsyuk:Logan}Logan J.D., Invariant variational principles, Academic Press, New York, 1977.

\bibitem{matsyuk:MKaw}
        Kawaguchi~M., An introduction to the theory of higher order spaces. II.~Higher order spaces in multiple parameters, {\it RAAG Memoirs} {\bf 4} (1968),
        578--592. 

\bibitem{matsyuk:DGA8}Matsyuk~R.Ya., Autoparallel variational
        description of the free relativistic top third order dynamics,
    in Proceedings of Eight International Conference
    ``Dif\/ferential Geometry and Its Applications''
    (August~27--31, 2001, Opava, Czech Republic),
    Editors O.~Kowalski~et al., Silesian Univ., Opava, 2002,
    447--452.

\bibitem{matsyuk:de Leon}de L\'eon M., Rodrigues P. R., Generalized
        classical mechanics and f\/ield theory, Elsevier, Amsterdam, 1985.

\bibitem{matsyuk:MathMet13}Matsyuk R.Ya., On the existence of a Lagrangian
        for a system of ordinary dif\/ferential equations, {\it  Mat. Metody Fiz.-Mekh. Polya\/}
                {\bf 13} (1981), 34--38, 113 (in Russian).

\bibitem{matsyuk:DAN}Matsyuk R.Ya., Lagrangian analysis of invariant third-order equations of motion in relativistic classical particle mechanics,
        {\it Dokl. Akad. Nauk SSSR\/} {\bf 282} (1985), 841--844
        (English transl.: {\it Soviet Phys. Dokl.} {\bf 30} (1985), 458--460).

\bibitem{matsyuk:Arreaga}Arreaga G., Capovilla R., Guven J., Frenet--Serret dynamics,
        {\it Classical Quantum Gravity} {\bf 18} (2001), 5065--5083, \href{http://arxiv.org/abs/hep-th/0105040}{hep-th/0105040}.

\bibitem{matsyuk:Boundary Problems}Matsyuk R.Ya., The variational principle for geodesic circles,
        in Boundary Value Problems of Mathematical Physics,  Naukova Dumka, Kiev, 1981, 79--81  (in Russian).

\bibitem{matsyuk:Acatrinei}Acatrinei C.S., A path integral leading to higher order
        Lagrangians, {\it J. Phys. A: Math. Gen.} {\bf 40} (2007), F929--F933, \href{http://arxiv.org/abs/0708.4351}{arXiv:0708.4351}.

\bibitem{matsyuk:Dixon}
        Dixon W.G., Dynamics of extended bodies in general relativity~I. Momentum
        and angular momentum,
        {\it Proc. Roy. Soc. London. Ser.~A.} {\bf 314} (1970), 499--527.

\bibitem{matsyuk:YanoIshihara}Yano~K., Ishihara Sh., Tangent and cotangent
        bundles, Marcel Dekker, New York, 1973.

\bibitem{matsyuk:Ehresmann}Ehresmann Ch., Les prolongements d\'une vari\'et\'e
        dif\/f\'erentiable~I.
        Calcul des jets, prolongement principal,
        {\it C.~R.~Math. Acad. Sci. Paris} {\bf 233} (1951), 598--600.

\bibitem{matsyuk:OKrup}Kruprov\'a O., The geometry of ordinary
        variational equations, {\it Lect. Notes in Math.}, Vol.~1678,
        Springer-Verlag, Berlin, 1997.

\bibitem{matsyuk:Tulcz}Tulczyjew W.M., Sur la dif\/f\'erentielle de
        Lagrange, {\it C. R. Math. Acad. Sci. Paris} {\bf 280} (1975),
        1295--1298.

\end{thebibliography}
\end{document}